\documentclass[reprint,
superscriptaddress,
 amsmath,
 amssymb,
 aps,
]{revtex4-2}

\usepackage[utf8]{inputenc}

\usepackage{graphicx}
\usepackage{dcolumn}
\usepackage{bm}
\usepackage[colorlinks]{hyperref}
\hypersetup{
    linkcolor = blue,   
    citecolor = blue,      
    urlcolor  = blue    
}
\usepackage{color}
\usepackage{amsmath}

\begin{document}

\title{Dynamics of interacting bosons in a two-leg ring ladder with artificial magnetic flux and ac-driven modulations}

\author{L. Q. Lai}
\email{lqlai@njupt.edu.cn}
\affiliation{School of Science, Nanjing University of Posts and Telecommunications, Nanjing 210023, China}

\date{\today}

\begin{abstract}

We investigate the nonequilibrium dynamics of interacting bosons in a two-leg ring ladder pierced by an artificial magnetic flux, where the particles are initially localized in the central sites of both rings, and the ac-driven local energy shifts are applied to the remaining lattice sites. Within the mean-field approximation, we demonstrate the emergence of nonlinear self-trapping for strong interparticle interactions, and characterize the distinct excitation regimes in the absence of the inter-ring tunneling. The artificial magnetic flux typically introduces the Peierls phase factors, which induces complex-valued hopping amplitudes and leads to directed net particle currents along the chains. By further incorporating the finite inter-ring coupling and biased intra-ring hopping, we reveal that the tuning of the drive frequency and Peierls phase allows the precise control over both the intensity and direction of particle currents, which facilitates the transition between chiral and antichiral dynamics. These findings offer insights into the coherent manipulation of matter-wave transports in closed-loop lattice configurations and the exploration of nonequilibrium synthetic quantum systems in related fields.

\end{abstract}

\maketitle

\section{Introduction}

The generation of synthetic gauge fields has facilitated the coherent manipulation of nonequilibrium many-body dynamics \cite{cooper}, and emerged as a powerful tool for probing novel phases of quantum matter and underlying the description of a rich variety of physical systems, ranging from photonics and topological insulators \cite{ozawa,qi} to arrays of Rydberg atoms and ultracold quantum gases in optical lattices \cite{scholl,eckardt}. In these systems, the fundamental interplay between the artificial magnetic field, the interparticle interactions and the dimensionality drives exotic quantum phenomena, motivating intense theoretical and experimental investigations.

Interacting bosons on a linear ladder in the presence of a uniform magnetic filed has been the topic of previous theoretical works \cite{dhar,mueller,tokuno,tschischik,keles,orignac1,orignac2,zheng,sachdeva,piraud,orignac3,orignac4,greschner,buser,barbiero,xue1,xue2,xue3,zhang,citro,halati,mishra1,mishra2,mishra3}, as it is pretty straightforward to study the dynamical responses of particles to magnetic fields, and offer insights into two-dimensional many-body quantum systems. The artificial magnetic flux typically induces phase transitions when the strength of hopping across the ladder and the interactions between the particles are varied, and the specific geometry can be experimentally implemented with ultracold atoms in tunable optical lattices \cite{lin,atala,mancini,an,aidelsburger0,jia,yan,aidelsburger}, where the chiral dynamics has been observed under precisely adjustable flux and interactions. In addition, it has been demonstrated that the on-site gain and loss in a noninteracting ladder system with the flux-induced asymmetric transport and inter-leg coupling generally results in the antichiral dynamics, where particles in both legs of the ladder propagate unidirectionally in the same direction  \cite{wu,ye}.

\begin{figure}[htbp]
\includegraphics[width=0.85\columnwidth]{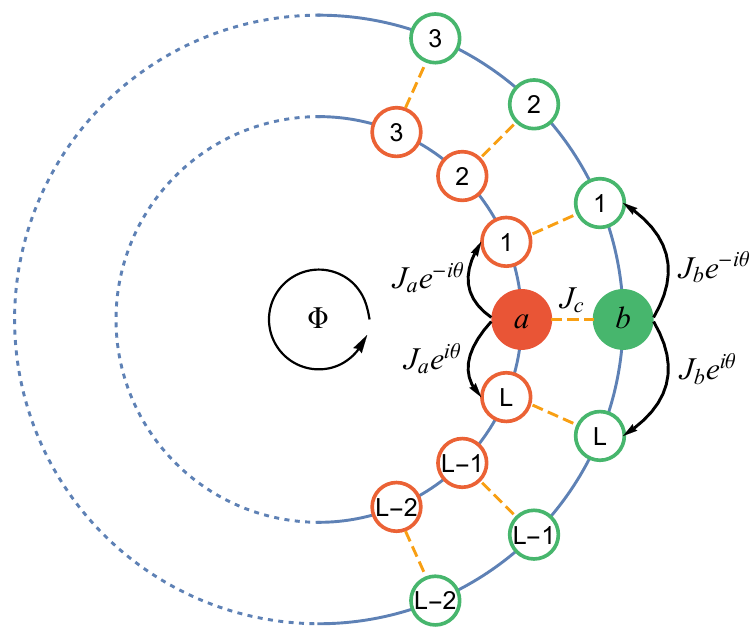}
\caption{Sketch of the two-leg ring ladder with an artificial magnetic flux denoted by $\Phi$. The red solid circle and the green solid circle represent the central sites $a$ and $b$, respectively, at which the particles are initially localized. The empty circles labeled by integers $1,2,3,\ldots,L$ correspond to the sites distributed along each lattice chain. The inter-ring coupling strength is characterized by the amplitude $J_{c}$, and the intra-ring hopping from site $\ell$ to $\ell+1$ has matrix element $J_{\nu}e^{-i\theta}$, while the one from $\ell+1$ to $\ell$ is $J_{\nu}e^{i\theta}$, with $\nu=\{a,b\}$. }
\label{ringlattice}
\end{figure}

There has also been increasing research interest in ultracold quantum gases subjected to the ring-shaped optical lattices \cite{amico1,amico2,amico8,minguzzi0,amico4,guilleumas1,amico5,molina,lai} and coupled rings \cite{oliinyk,bland1,bland2,nicolau,amico6,borysenko,schubert} due to the promising applications in future atomic circuits, and a finite-length two-leg ladder can readily be mapped to a two-leg ring ladder. Related intriguing physics in this specific configuration has been recently explored  \cite{richaud,minguzzi1,minguzzi2,minguzzi3,guilleumas2,guilleumas3}, while most of the efforts focused on the undriven quantum dynamics or ladders with homogeneous interacting particles on lattice sites. Time-periodic modulation serves as a versatile tool for engineering quantum states and controlling the nonequilibrium dynamics that do not appear in static systems \cite{eckardt}, and a combination of the artificial magnetic flux with a joint modulation of both on-site energies and tunneling parameters allows further investigations on persistent currents and quantum simulations in dimensions larger than one. Moreover, the population imbalance between the two rings generates an additional phase, where one may also reorient the direction of the particle currents in the context of an ac driving and biased intra-ring hopping \cite{mueller,molina}. In this work, we introduce a two-leg ring ladder pierced by an artificial magnetic flux, where the bosons are initially prepared in the central sites. We apply the ac-driven local energy shifts to the remaining lattice sites, and numerically analyze the excitation regimes for different interactions. We also consider biased intra-ring hopping amplitudes, and modulate the drive frequency and Peierls phase to reveal the chiral and antichiral dynamics.

The paper is organized as follows. In Sec.~\ref{model}, we introduce the two-leg ring ladder model, and describe the mean-field approach and the corresponding equations of motion. In Sec.~\ref{result}, we present the numerical analysis and related discussions for various dynamical properties. We summarize the results in Sec.~\ref{summary}.

\section{Theoretical Model}\label{model}

We begin by considering a two-leg ring ladder system pierced by an artificial magnetic flux, as schematically shown in Fig.~\ref{ringlattice}, where the bosonic atoms are initially prepared in the central sites $a$ and $b$, and the remaining lattice sites experience ac-driven local energy shifts. In the tight-binding limit, the system can thus be mathematically described by the Hamiltonian
\begin{eqnarray} \label{hamiltonian}
\hat{H} &=& \frac{U}{2}\sum_{\ell=0}^{L}\left(\hat{a}_{\ell}^{\dag}\hat{a}_{\ell}^{\dag}\hat{a}_{\ell}\hat{a}_{\ell}+\hat{b}_{\ell}^{\dag}\hat{b}_{\ell}^{\dag}\hat{b}_{\ell}\hat{b}_{\ell}\right) \nonumber \\
&&+F\left(t\right) \sum_{\ell=1}^{L}\left(\hat{a}_{\ell}^{\dag}\hat{a}_{\ell}+\hat{b}_{\ell}^{\dag}\hat{b}_{\ell}\right)-J_{c}\sum_{\ell=0}^{L}\left(\hat{a}_{\ell}^{\dag}\hat{b}_{\ell}+{\rm H.c.}\right) \nonumber \\
&&-\sum_{\ell=0}^{L}\left(J_{a}e^{-i\theta}\hat{a}_{\ell+1}^{\dag}\hat{a}_{\ell}+J_{b}e^{-i\theta}\hat{b}_{\ell+1}^{\dag}\hat{b}_{\ell}+{\rm H.c.}\right).
\end{eqnarray}
Here, $\hat{a}^{\dag}_{\ell}$ $(\hat{a}_{\ell})$ and $\hat{b}^{\dag}_{\ell}$ $(\hat{b}_{\ell})$ are the bosonic creation (annihilation) operators at site $\ell$ in each ring, and $U$ characterizes the on-site pairwise interaction. The time-dependent modulation $F(t)=\mu+M\cos(\omega t)$ represents the local energy shift applied to sites $\ell\geq 1$, where $\mu$ is a constant term and $M\cos(\omega t)$ is a cosinoidally oscillating component with $M$ being the drive strength and $\omega$ being the drive frequency, which can be specifically implemented in an experiment by using a spatial optical modulator \cite{mcgloin,molina}. The intra-ring hoppings between nearest-neighboring sites are quantified by complex amplitudes $J_{\nu}e^{\pm i\theta}$, with $\nu=\{a,b\}$ and $\theta$ being the Peierls phase resulting from the artificial magnetic field, and the inter-ring coupling strength is denoted by $J_{c}$.

In the mean-field approximation, the field operators can be reasonably replaced by their expectation values $\nu_{\ell}=\langle\hat{\nu}_{\ell}\rangle\equiv\varphi_{\nu,\ell}$, where physically $|\varphi_{\nu,\ell}|^2$ corresponds to the number of particles on site $\ell$ of ring $\nu$. Accordingly, the Heisenberg equations of motion are straightforward ($\hbar=1$ throughout this paper)
\begin{eqnarray}
i\frac{\partial}{\partial t}
\left[
\begin{array}{cc}
    \varphi_{a,\ell} \\
    \varphi_{b,\ell}
\end{array}
\right] &=&
U
\left[
\begin{array}{cc}
    |\varphi_{a,\ell}|^{2}\varphi_{a,\ell}  \\
    |\varphi_{b,\ell}|^{2}\varphi_{b,\ell} 
\end{array}
\right] -
J_{c}
\left[
\begin{array}{cc}
    \varphi_{b,\ell}  \\
    \varphi_{a,\ell} 
\end{array}
\right] \nonumber \\
&&-
\left[
\begin{array}{cc}
     J_{a}e^{i\theta}\varphi_{a,\ell+1} \\
     J_{b}e^{i\theta}\varphi_{b,\ell+1}
\end{array}
\right] -
\left[
\begin{array}{cc}
     J_{a}e^{-i\theta}\varphi_{a,\ell-1} \\
     J_{b}e^{-i\theta}\varphi_{b,\ell-1}
\end{array}
\right] \nonumber \\
&&+F(t)
\left[
\begin{array}{cc}
    \varphi_{a,\ell}\delta_{\ell,m} \\
    \varphi_{b,\ell}\delta_{\ell,m}
\end{array}
\right],
\end{eqnarray}
where the integer $m=1,2,3,\ldots$ indicates the location of sites with the energy shifts. In what follows, we focus on a fixed lattice chain of length $101$ with the central site being indexed by ``0" and $L=100$, where the main dynamical features apply to other finite lengths, and employ the periodic boundary condition to identify $\varphi_{\nu,L+1}$ with $\varphi_{\nu,0}$ and $\varphi_{\nu,-1}$ with $\varphi_{\nu,L}$, respectively. Without any loss of generality, in the numerics we take $J_{a}=1$ as the energy unit, such that the times are measured in units of $\hbar/J_{a}$. Since the particles are initially localized in the central sites of both rings, we also set the initial values for the stationary states as $\varphi_{\nu,\ell}\left(t<0\right)=\delta_{\ell,0}$, and explore the nonlinear dynamics of the system by introducing the interaction effects, the local energy shifts, and the biased hopping amplitudes.

\section{Results} \label{result}

\subsection{Self-trapping dynamics and excitation regimes}

In this section, we consider the situation with the intra-ring hopping amplitude $J_{a}=J_{b}$ while the inter-ring coupling strength is $J_{c}=0$, such that the system is divided into two independent ring lattices without particle exchange. Under this circumstance, they share identical dynamics in the presence of the same artificial magnetic flux and ac-driven local energy shifts, and it is convenient to demonstrate the excitation regimes.

\begin{figure}[htbp]
    \includegraphics[width=\columnwidth]{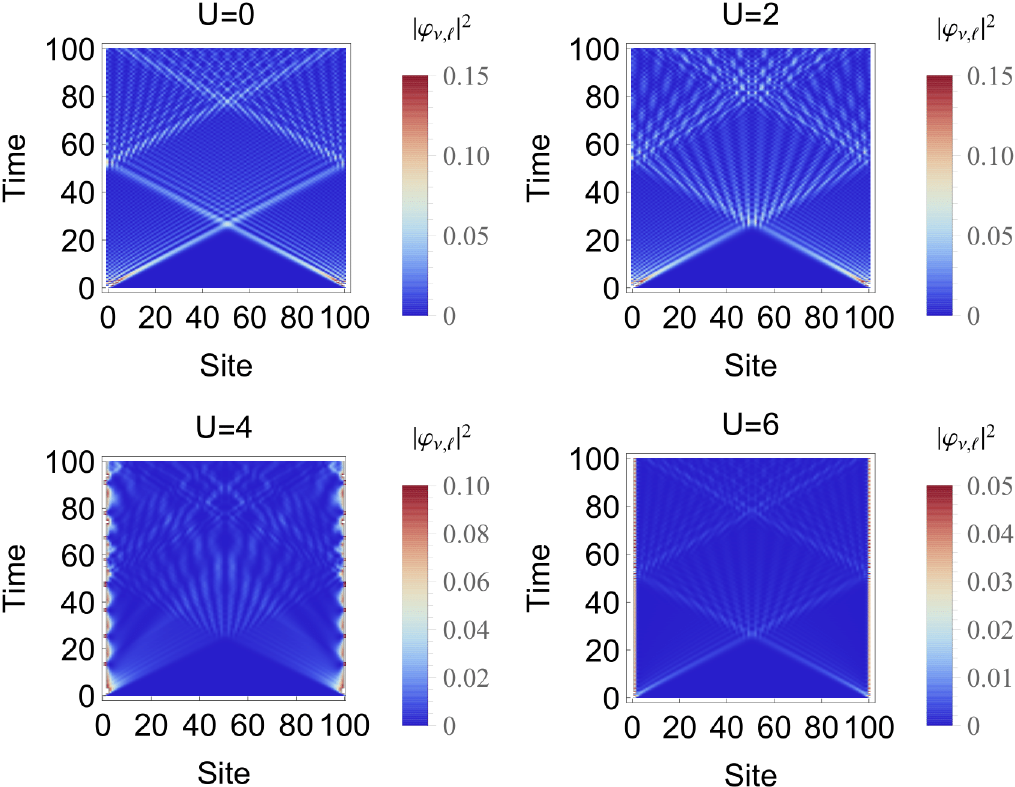}
    \caption{Time evolution of the number of particles on the $\ell$th site of ring $\nu$ in the presence of different interaction strengths $U=0$, $2$, $4$ and $6$, respectively, with the Peierls phase $\theta=0$, the inter-ring coupling strength $J_{c}=0$, and the intra-ring hopping amplitude $J_{\nu}=1$. The periodic boundary condition is employed, and the local energy shifts are absent.}
    \label{jetarrayUtot}
\end{figure}

We first analyze the nonlinear interaction effects on the time evolution of particles, as shown in Fig.~\ref{jetarrayUtot}. One can find distinct dynamical evolutions depending on the interaction strength $U$: in the noninteracting case ($U=0$), the particles can readily escape from the central site and counterpropagate freely along the chain, resembling a back-and-forth trace at long times with occasional occupation on the lattice sites. At weak interactions (exemplified by $U=2$), the free evolution is slightly weakened, where the particles exhibit enhanced occupation at specific sites over time. As for stronger interactions (e.g., $U=4$ and $U=6$), however, the propagation is significantly suppressed, resulting in the disorder-free localization phenomenon, namely, the self-trapping effect, where the particles remain predominantly confined in the central site.

\begin{figure}[htbp]
    \includegraphics[width=\columnwidth]{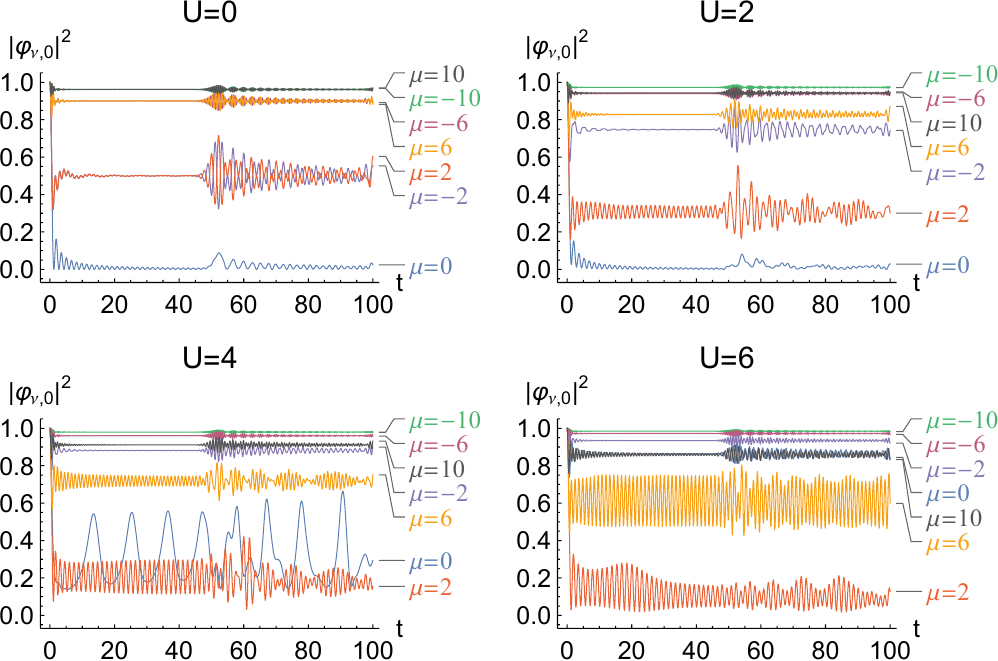}
    \caption{Decay of particles in the central site for different dc amplitude $\mu$, when the ac driving is absent. Here, the interaction strengths are $U=0$, $2$, $4$ and $6$, respectively, and the Peierls phase is $\theta=0$. The inter-ring coupling strength is $J_{c}=0$, and the intra-ring hopping amplitude is $J_{\nu}=1$.}
    \label{totaoMu}
\end{figure}

The modulation of the local energies with respect to sites $\ell\geq 1$ can be generally regarded as a substrate potential that makes the central site vary from the others, and here we consider the case with the intensity of the shift being constant, i.e., the amplitude of the ac-driven component is $M=0$. As can be plainly seen in Fig.~\ref{totaoMu}, for generic interaction strengths a negative dc amplitude $\mu$ typically strengthens the self-trapping, as it increases the energy offset between the central site and the surrounding sites, thereby the particle number of the central site simply oscillates and the atoms remain bound. Conversely, for moderate positive amplitudes the decay of the particles can be enhanced, with stronger interactions leading to more prominent enhancement and oscillations, since in this case the central site may be intuitively considered as a potential well, and 
the effective potential difference holds modest. In addition, for sufficiently large positive amplitudes $\mu\gg J_{\nu}$ (e.g., $\mu=10$), particle localization can also occur. Note that at $\mu=0$ the system exhibits free evolution, where the decay is the most significant for weak interactions but greatly hindered with rapid oscillation for stronger interactions, which is consistent with Fig.~\ref{jetarrayUtot}. In what follows, we primarily take the dc amplitude $\mu=10$ into account, whence the supercurrents are virtually suppressed before the ac-driven modulation is turned on \cite{molina}, such that we can explicitly clarify the ac-driven scenario.

\begin{figure}[htbp]
    \includegraphics[width=\columnwidth]{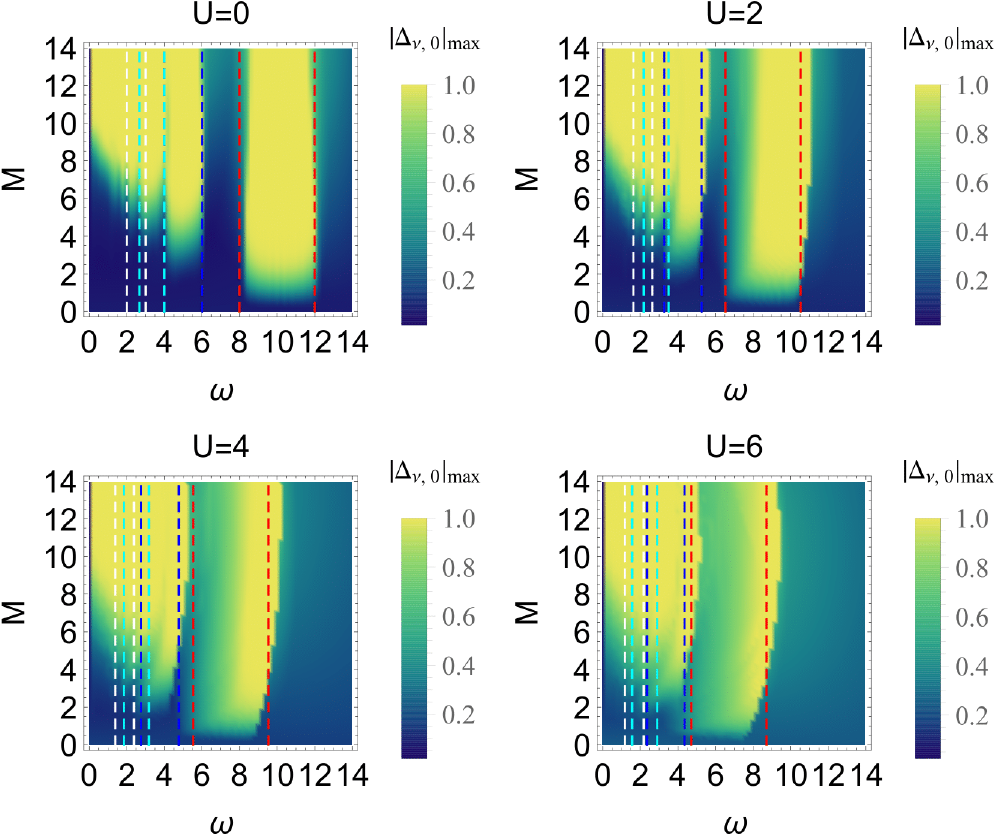}
    \caption{The maximum particle number difference $\vert\Delta_{\nu,0}\vert_{\rm{max}}$ vs drive strength $M$ and  drive frequency $\omega$ under different interaction strength $U=0$, $2$, $4$, and $6$, respectively. Here, we have taken the dc amplitude $\mu=10$, the intra-ring hopping amplitude $J_{\nu}=1$, the inter-ring coupling strength $J_{c}=0$, and the Peierls phase $\theta=0$. The simulation time is $t_{s}=50$, and the dashed lines delineate the regimes $\omega_{\pm}=\left(\pm2J_{\nu}+\Delta\epsilon\right)/n$ corresponding to the $n$th-order excitations.}
    \label{dtota0Momega}
\end{figure}

With the application of the ac driving, the local energies of sites $\ell\geq 1$ are shifted periodically in time. One can thus define the difference of the particle number
\begin{eqnarray}
\Delta_{\nu,0}\left(t\right)=\vert \varphi_{\nu,0}\left(t\right) \vert^{2}-\vert \varphi_{\nu,0}\left(t=0\right)\vert^{2}.
\end{eqnarray}
Here we are mainly interested in the short-time behaviors before the outgoing particles return to the central site, so we perform the simulation over $0<t_{s}<50$, and calculate the maximum difference $\vert \Delta_{\nu,0}\vert_{\rm{max}}=\underset{t\in t_{s}}{\rm{max}}\vert\Delta_{\nu,0}\left(t\right)\vert$. Figure~\ref{dtota0Momega} illustrates how the maximum difference depends on the drive strength $M$ and the drive frequency $\omega$. For generic drive frequencies and weak ac-driven modulation amplitude, the system can be quite stable and only a tiny portion of particles are excited. We see that for roughly $-2J_{\nu}<\omega-\Delta\epsilon<2J_{\nu}$, with $\Delta\epsilon$ denoting the energy gap between the central site and the nearest-neighboring sites, even a small $M$ leads to particle excitation. This is related to the energy conservation: In the noninteracting limit $U=0$ and in the absence of the Peierls phase and ac modulation, one can directly perform the Fourier transform on Hamiltonian~(\ref{hamiltonian}) to obtain the Bloch Hamiltonian $H\left( k\right)$ in the momentum space and figure out the dispersion relation $\epsilon_{k}=-2J_{\nu}\cos\left(k\right)$, where in our case the gap is explicitly the dc amplitude $\mu$, requiring the drive frequency $-2J_{\nu}+\mu<\omega<2J_{\nu}+\mu$ for exciting the particles. At weak interactions, in combination with the Bogoliubov approximation \cite{plischke} gives the gap
\begin{eqnarray}
\Delta\epsilon=\mu+\epsilon_{k}-\sqrt{\epsilon_{k}^{2}+2U\epsilon_{k}},
\end{eqnarray}
where finite interaction strengths $U$ specifically renormalize the boundaries of the bands. There are further bands of instability with $-2J_{\nu}+\Delta\epsilon<n\omega<2J_{\nu}+\Delta\epsilon$ for integers $n$, as the drive also provides energy in multiples, and the $n$th-order bands overlap with each other. Within the allowed regions, the ac amplitude mainly affects the intensity of the tunneling, where a larger $M$ typically results in more excited particles with a larger $\vert \Delta_{\nu,0}\vert_{\rm{max}}$. Note that for the drive frequency $\omega\approx0$, the ac-driven component $M\cos\left(\omega t\right)$ can be readily absorbed into the dc amplitude $\mu$ as an even larger positive amplitude, and hence few particles escape from the central site.

\begin{figure}[htbp]
    \includegraphics[width=\columnwidth]{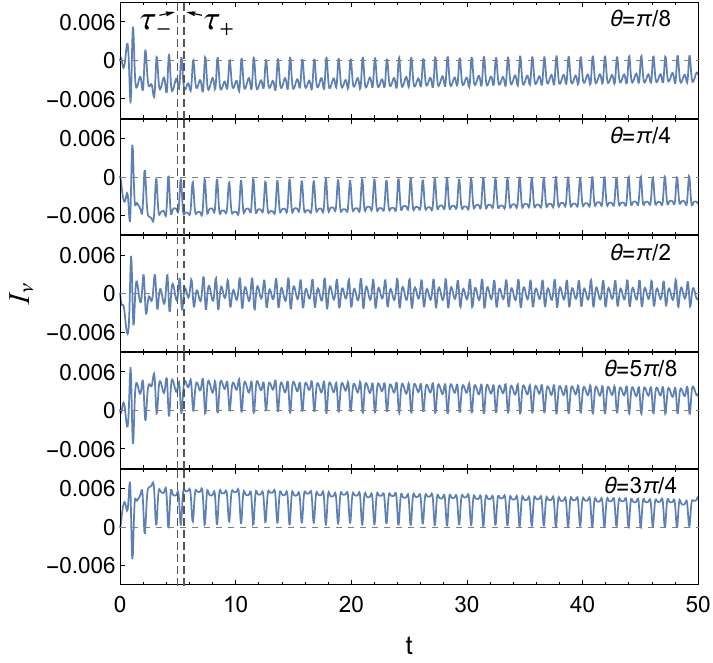}
    \caption{Time dependence of the net particle currents with Peierls phases $\theta=\pi/8$, $\pi/4$, $\pi/2$, $5\pi/8$, and $3\pi/4$, respectively. The dashed lines corresponding to $\tau_{-}=19\times \frac{\pi/2}{\omega}$ and $\tau_{+}=21\times \frac{\pi/2}{\omega}$ indicate an exemplified region of the typical oscillation of particle currents within half a driving cycle. Here, we have kept the interaction strength $U=6$ and the dc amplitude $\mu=10$, and taken the driving parameters $M=8$ and $\omega=6$. The intra-ring hopping amplitude is $J_{\nu}=1$, and the inter-ring coupling strength is $J_{c}=0$.}
    \label{pcurrentatot}
\end{figure}

The Peierls phase typically breaks the time-reversal symmetry, where the particles hop on the lattice with complex tunneling matrix elements. Basically, the particle current flowing from site $\ell$ to $\ell+1$ in ring $\nu$ yields
\begin{eqnarray}
I_{\nu,\ell}\left(t\right)=-iJ_{\nu}e^{-i\theta}\varphi_{\nu,\ell+1}^{*}\varphi_{\nu,\ell}+iJ_{\nu}e^{i\theta}\varphi_{\nu,\ell}^{*}\varphi_{\nu,\ell+1},
\end{eqnarray}
and can be experimentally determined through the phase gradient by using interferometry \cite{molina,pezze}. To gain insight into the directed transport process, we consider the parameters based on the regimes in Fig.~\ref{dtota0Momega}, and calculate the net particle currents from
\begin{eqnarray}
I_{\nu}\left(t\right)=\frac{1}{L}\sum_{\ell}{I}_{\nu,\ell}(t).  
\end{eqnarray}
As shown in Fig.~\ref{pcurrentatot}, when the ac driving is applied, the net particle currents for the phases $\theta=\pi/8$ and $\theta=\pi/4$ exhibit rapid oscillations around typical negative values (clockwise flows based on the sketch of Fig.~\ref{ringlattice}), with their amplitudes slightly damping with time. In sharp contrast, for $\theta=5\pi/8$ and $\theta=3\pi/4$ both currents reverse directions, oscillating around positive values while maintaining the same magnitudes as their counterparts ($\theta=\pi/8$ and $\theta=\pi/4$, respectively). This correspondence arises from the equivalencies of the exponential, where $e^{i\left(\alpha+\pi/2\right)}=ie^{i\alpha}$ ($\alpha$ is real) and the currents obey $I_{\nu}(t)_{\theta=\alpha}=-I_{\nu}(t)_{\theta=\alpha+\pi/2}$, and at the specific phase $\theta=\pi/2$ the current appears to oscillate around zero. As a consequence, the direction of $I_{\nu}$ can be simply varied by either inverting the Peierls phase from $\theta$ to $-\theta$, or making use of the phase-dependent symmetry.

There are also two exemplified time scales $\tau_{-}$ and $\tau_{+}$ illustrated in Fig.~\ref{pcurrentatot}. Since the ac driving follows a cosinusoidal oscillation, the total amplitude stays even larger than the fixed dc amplitude during the typical half cycle ($t\in\left[\tau_{-},\tau_{+}\right]$), which is followed by a reciprocal tendency, i.e., its intensity first increases and then decreases, leading to the revival of the particle current. In the other half of the driving cycle, however, the total amplitude keeps smaller than the dc amplitude, which stimulates the excitations with a relatively smoother oscillation, and results in a comblike pattern overall. Notably, at sufficiently long times the particle occupation on the lattice sites eventually saturates, and the system reaches a dynamic equilibrium, where the currents along the chain remain small but nearly persistent.

\begin{figure}[t]
    \includegraphics[width=\columnwidth]{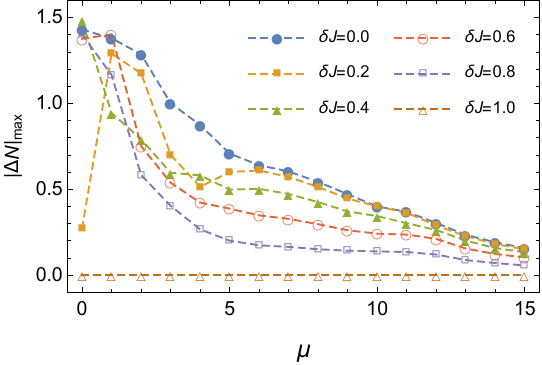}
    \caption{The maximum particle imbalance $\vert\Delta N\vert_{\rm{max}}$ as a function of the dc amplitude $\mu$ for different hopping amplitude $J_{b}$. Here, the inter-ring coupling strength is $J_{c}=1$, and the interaction strength is $U=6$. The ac driving is absent, and the calculation is performed with $0<t_{s}<50$.}
    \label{dnanbMu}
\end{figure}

\subsection{Chiral and antichiral dynamics with biased intra-ring hopping}

We now turn to study the dynamics when the inter-ring coupling strength $J_{c}$ is nonzero, where the corresponding lattices are coupled and the system changes into a two-leg ring ladder. For convenience, we keep $J_{c}=1$, and still lay emphasis on the interaction strength $U=6$ in order to maintain consistency with the former case. The particle imbalance between the rings can be defined as
\begin{eqnarray}
\Delta N\left(t\right)=N_{a}\left(t\right)-N_{b}\left(t\right),
\end{eqnarray}
where
\begin{eqnarray}
N_{\nu}\left(t\right)=\sum_{\ell=0}^{L}\vert \varphi_{\nu,\ell}\left(t\right)\vert^{2}
\end{eqnarray}
is the total particle number in ring $\nu$, and the maximum imbalance is computed by $\vert \Delta N\vert_{\rm{max}}=\underset{t\in t_{s}}{\rm{max}}\vert\Delta N\left(t\right)\vert$. We also denote $\delta J=J_{b}/J_{a}$, where the biased hopping is incorporated by taking $\delta J \neq 1$. As shown in Fig.~\ref{dnanbMu}, for generic $\delta J$ the maximum imbalance $\vert \Delta N\vert_{\rm{max}}$ appears to decrease exponentially with the dc amplitude $\mu$, as the particles tend to become localized for large positive amplitudes and the imbalance is reduced. At a fixed dc amplitude $\mu$, larger $\delta J$ values result in smaller imbalances, and for the typical $\delta J=1$ the rings remain balanced with near-zero particle imbalance, as expected.

\begin{figure}[htbp]
    \includegraphics[width=\columnwidth]{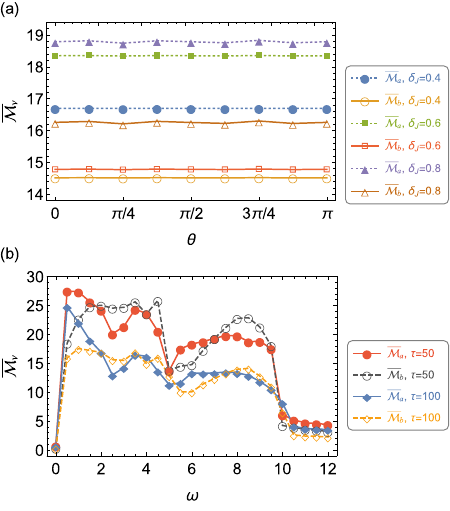}
    \caption{Time-averaged center of mass in ring $\nu$. (a) $\overline{\mathcal{M}}_{\nu}$ vs $\theta$ for different bias $\delta J=0.4$, $0.6$, and $0.8$, respectively, with the drive frequency $\omega=6$ and the driving interval $\tau=50$. (b) $\overline{\mathcal{M}}_{\nu}$ vs $\omega$ with the bias $\delta J=0.6$ and Peierls phase $\theta=\pi/4$ under specific driving intervals $\tau=50$ and $100$, respectively. The other parameters are $J_{c}=1$, $U=6$, $\mu=10$, and $M=8$.}
    \label{comtot}
\end{figure}

Due to the presence of the biased hopping, the drive frequencies that meet the excitation conditions may slightly differ between the two rings, especially when the Peierls phase is involved. Nevertheless, for finite interactions the bands overlap, which explicitly renormalizes the spectral boundaries and broadens the discrete bands into near-continuous ones (e.g., $U=4$ and $U=6$ in Fig.~\ref{dtota0Momega}), leading to distinct particle excitations in the vicinity of resonant frequencies. We thus keep the dc amplitude as $\mu=10$, and proceed to calculate the time-averaged center of mass of ring $\nu$
\begin{eqnarray}
\overline{\mathcal{M}}_{\nu}=\frac{1}{\tau}\int_{0}^{\tau}\sum_{\ell}\ell\vert\varphi_{\nu,\ell}\left(t\right)\vert^{2}.
\end{eqnarray}
We begin by analyzing the center of mass $\overline{\mathcal{M}}_{\nu}$ as a function of $\theta$ for different $\delta J$, and clarify the corresponding particle occupation, as shown in Fig.~\ref{comtot}(a). In general, the magnetic flux appears to have little impact on $\overline{\mathcal{M}}_{\nu}$, inducing only slight oscillations. $\overline{\mathcal{M}}_{a}$ remains larger than $\overline{\mathcal{M}}_{b}$, as the particles in ring $a$ hop faster than those in ring $b$ due to the biased hopping $\delta J<1$, leading to the accumulation at larger site indices of ring $a$. As the bias increases, $\overline{\mathcal{M}}_{\nu}$ tends to rise accordingly, which in turn reduces the difference between $\overline{\mathcal{M}}_{a}$ and amplifies the difference between $\overline{\mathcal{M}}_{b}$, respectively.

The influence of the drive frequency $\omega$ on $\overline{\mathcal{M}}_{\nu}$ can also be examined, as illustrated in Fig.~\ref{comtot}(b). For a driving interval of $\tau=50$, one can find $\overline{\mathcal{M}}_{a}\approx\overline{\mathcal{M}}_{b}\approx0$ at $\omega=0$, since most of the particles are still confined in the central sites of both rings under such conditions. At around $\omega=1$, $\overline{\mathcal M}_{\nu}$ peaks with $\overline{\mathcal{M}}_{a}>\overline{\mathcal{M}}_{b}$, and beyond this point they decrease while oscillating alternately with the increase of the drive frequency $\omega$, corresponding to the particle exchange between the rings. For $\omega>10$, they stabilize with $\overline{\mathcal{M}}_{a}\gtrsim\overline{\mathcal{M}}_{b}$ even when the drive frequency is further increased, which is consistent with the excitation regimes of the system. In the other case of $\tau=100$, the behaviors of $\overline{\mathcal{M}}_{a}$ and $\overline{\mathcal{M}}_{b}$ roughly mirror their counterparts at $\tau=50$, while maintaining smaller magnitudes overall. Note that in the ring-lattice geometry, particles undergo back-and-forth motions (c.f. Fig.~\ref{jetarrayUtot}), and for sufficiently long driving intervals the occupation on specific sites may gradually saturate, which differs from a two-leg ladder that the population stacks at one end manifesting the skin dynamics \cite{zhang}.

\begin{figure}[htbp]
    \includegraphics[width=\columnwidth]{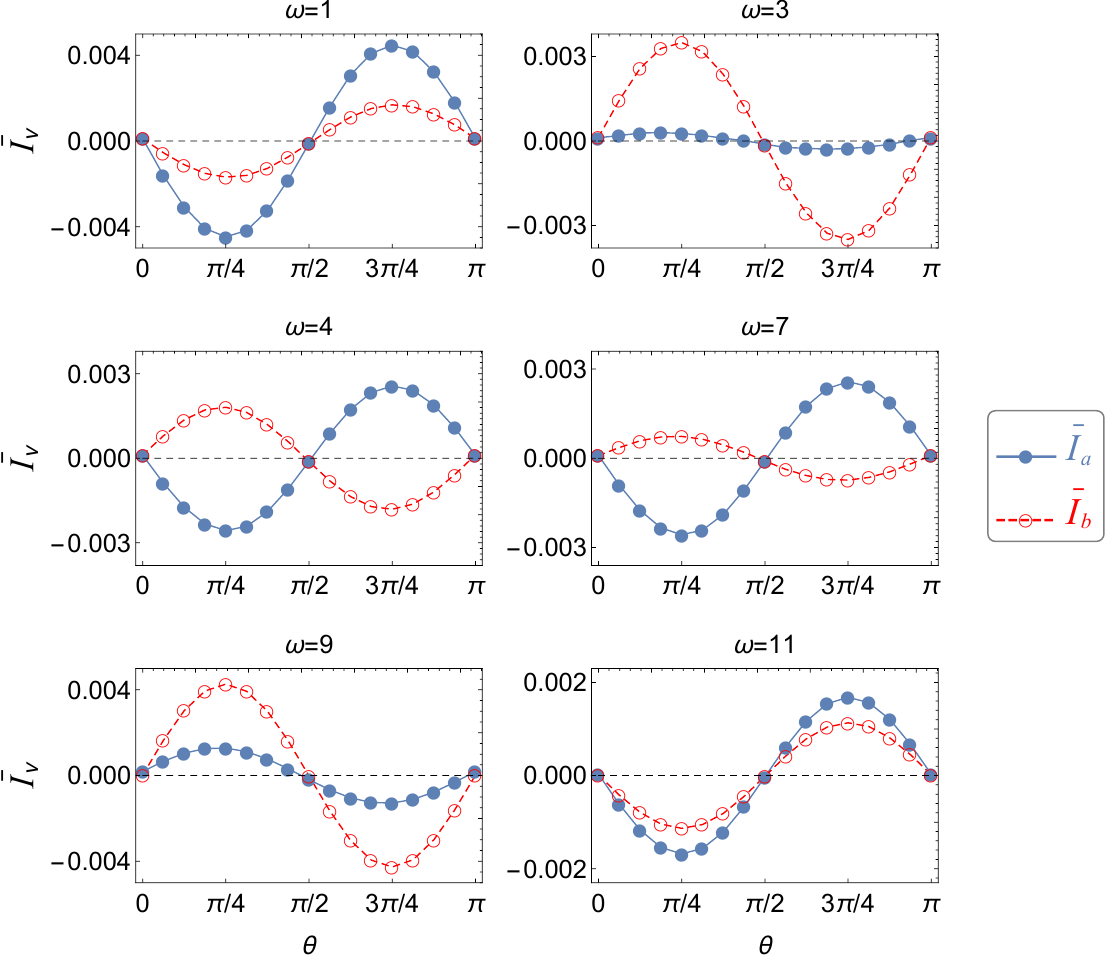}
    \caption{Asymptotic mean particle current $\bar{I}_{\nu}$ as a function of the Peierls phase $\theta$ under typical drive frequencies $\omega=1$, $3$, $4$, $7$, $9$ and $11$, respectively. Here, we have taken $J_{c}=1$, $\delta J=0.6$, $U=6$, $\mu=10$ and $M=8$, and the driving interval is kept as $\tau=50$.}
    \label{mpcurrenttot}
\end{figure}

We further take the moderate bias $\delta J=0.6$ as a typical example, and illustrate the dependence of the mean particle currents on the drive frequency $\omega$ and the Peierls phase $\theta$. The asymptotic mean particle current along ring $\nu$ can be calculated by $\bar{I}_{\nu}=\langle I_{\nu}\left(t\right)\rangle_{\tau}$, where $\langle\ldots\rangle_{\tau}$ represents the time average over a driving interval $\tau$. As can be seen in Fig.~\ref{mpcurrenttot}, at low frequency $\omega=1$ currents $\bar{I}_{a}$ and $\bar{I}_{b}$ maintain identical directions with different intensities, while they both reverse when the drive frequency is tuned to $\omega=3$, with one of the intensities being increased and the other being decreased. For the case of $\omega=4$, the two currents flow in opposite directions, and at $\omega=7$ this pattern continues, which roughly mirrors the behavior of $\omega=4$, and the intensity of $\bar{I}_{a}$ keeps almost unchanged and that of $\bar{I}_{b}$ is slightly reduced. The currents realign at higher frequencies $\omega=9$ and $\omega=11$, where they again flow in the same direction.

\begin{figure}[htbp]
    \includegraphics[width=\columnwidth]{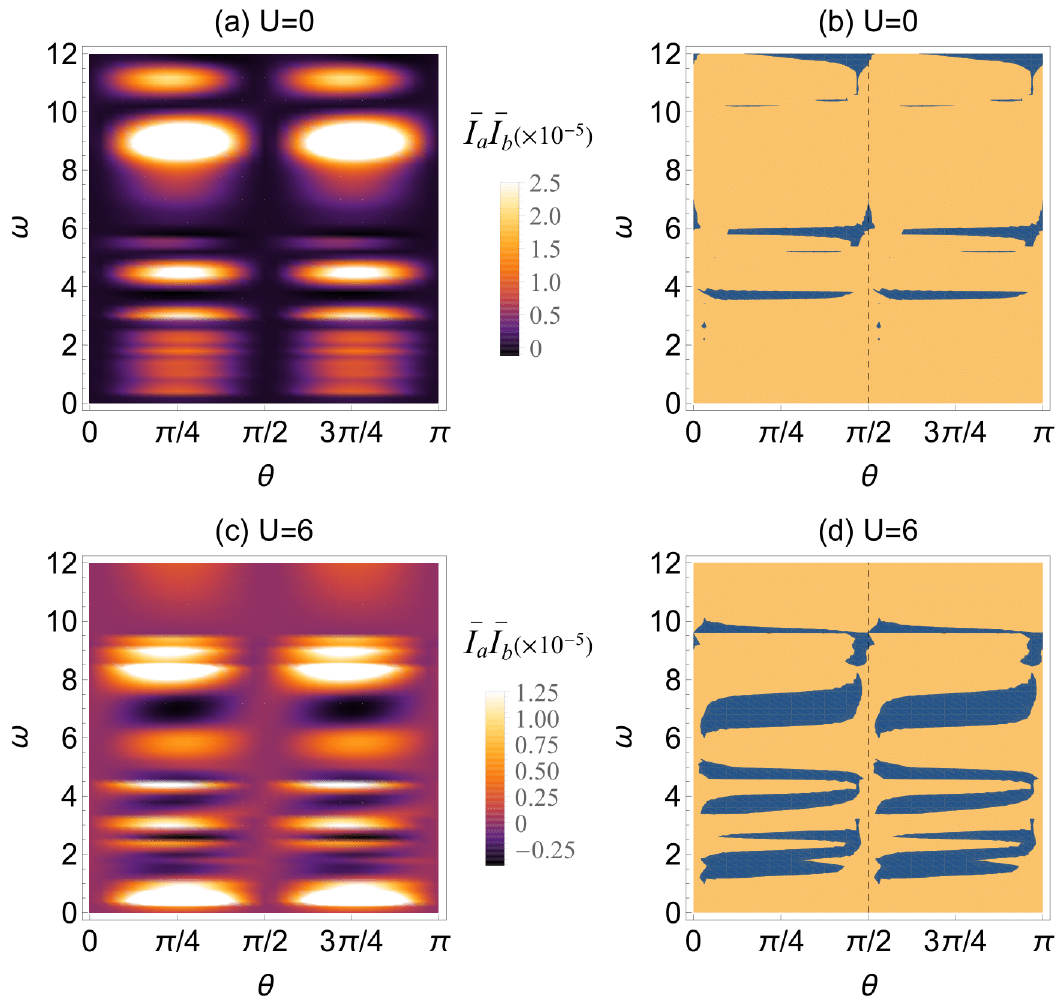}
    \caption{Regimes of the chiral and antichiral dynamics for interaction strength $U=0$ and $U=6$, respectively. (a) and (c) $\bar{I}_{a}\bar{I}_{b}$ vs the drive frequency $\omega$ and the Peierls phase $\theta$. (b) and (d) The phase diagram in the $\theta$-$\omega$ plane. The blue shading denotes the region for the chiral dynamics, while the orange shading corresponds to that of the antichiral dynamics. Here, we have taken $J_{c}=1$, $\delta J=0.6$, $\mu=10$, and $M=8$, and the driving interval is kept as $\tau=10$.}
    \label{mpctransition}
\end{figure}

To explore more quantitatively the chiral and antichiral dynamics, we identify the directions of the particle currents along the two rings  under different interactions, where they have opposite directions with $\bar{I}_{a}\bar{I}_{b}<0$ and the same direction with $\bar{I}_{a}\bar{I}_{b}>0$, respectively, which can be visible after short-time evolutions. One can see from Figs.~\ref{mpctransition}(a) and \ref{mpctransition}(c) that the introduction of the particle interaction and the ac-driven modulation notably shifts the bands with increased overlap regions, and also varies the relative directions of particle currents between the rings. For the interaction strength $U=6$, there are substantial regions corresponding to $\bar{I}_{a}\bar{I}_{b}<0$, where the chiral dynamics is exhibited, while at $U=0$ the value of $\bar{I}_{a}\bar{I}_{b}$ keeps predominantly positive, showcasing explicitly the antichiral dynamics. The phase $\theta=\pi/2$ serves as the cental line for the excitations according to the exponential equivalencies, with the features repeating on both sides. We characterize the chiral (blue shading) and antichiral (orange shading) dynamics in the $\theta$-$\omega$ plane by analyzing the sign of $\bar{I}_{a}\bar{I}_{b}$ in Figs.~\ref{mpctransition}(b) and \ref{mpctransition}(d), where $\bar{I}_{a}\bar{I}_{b}\approx0$ marks the chiral-antichiral transition boundaries. In the noninteracting case, the chirality appears at well-defined, discrete narrow bands in frequency $\omega$, and the spectrum is antisymmetric around $\theta=\pi/2$, while at $U=6$ the bands are significantly enlarged and fragmented for broader ranges of frequency, and thus the antichiral regions relatively contract. This suggests that chirality survives strong interactions, and their deformation signals a transition between the chiral and antichiral dynamics, which disappears ultimately due to the self-trapping effect. Accordingly, the intensity and direction of particle currents can be systematically controlled by manipulating the chiral-anticiral transition through joint tunings of the drive frequency and the Peierls phase.

\section{Summary and Outlook}\label{summary}

We have introduced a two-leg ring ladder pierced by an artificial magnetic flux, and explored the nonequilibrium dynamics of interacting bosons in the presence of ac-driven local energy shifts. The particles are initially localized in the central sites of both rings, while the local energy shifts are applied to the remaining lattice sites. Strong interparticle interactions lead to nonlinear self-trapping and confine particles in the initial sites, while weak interactions permit free propagation across the chains. Introducing a static energy offset to the surrounding sites allows further control over the particle localization, with either enhancement or suppression depending on the magnitude of the dc amplitude.

The application of an ac driving induces particle excitations, where the distinct spectra are determined by the interplay between drive frequency, interaction strength, and local energy gap. When a finite inter-ring coupling is present and the intra-ring hopping is biased, the system exhibits chiral (currents flow in opposite directions) and antichiral dynamics (currents flow in the same direction), and the intensity and direction of particle currents can be precisely determined by jointly modulating the drive frequency, Peierls phase and interaction strength, which stimulates the chiral-antichiral transition.

This specific geometry can be experimentally realized with ultracold bosonic atoms in ring-shaped optical lattices, with the tunable tunneling and on-site energies being implemented via spatial optical modulators and artificial gauge fields \cite{molina,mcgloin}. A weak local trap may be applied to the central sites of the rings before the atoms are loaded, such that they would be initially localized, and one subsequently switches off the trap, releasing the particles to evolve under the ac driving and modulating the interactions via Feshbach resonance. One could naturally extend the system to more complex configurations, such as multileg ring networks and two-dimensional lattices, and explore richer dynamics including topological pumping, quantum Hall-like behaviors and superfluid qubit systems \cite{amico4,amico7}. Moreover, the precise control of persistent currents and the transition between chiral and antichiral transports may contribute to quantum information transfer \cite{amico4,amico5}, novel atomtronic circuits \cite{amico9} and topological quantum simulations \cite{davoodi1,davoodi2}.

\section*{Acknowledgments}
This work was supported by National Natural Science Foundation of China (Grant No.~12505022) and the Natural Science Research Start-up Foundation of Recruiting Talents of Nanjing University of Posts and Telecommunications (Grant No.~NY223065).

\end{document}